\documentclass[twocolumn]{aastex63}

\usepackage{xcolor} 

\definecolor{mygreen}{RGB}{0, 185, 118}

\usepackage{soul}

\usepackage{booktabs}
\usepackage{amsmath, amssymb, amsfonts, latexsym}        
\usepackage{graphicx}   
\usepackage{blindtext}
\usepackage{enumitem}
\DeclareMathSizes{10}{10}{7}{6}
\shorttitle{Coronal holes influence on CME deflections}
\shortauthors{Sahade et al.}
\graphicspath{{./}{figures/}}

\begin{document}

\title{Influence of coronal holes on CME deflections: numerical study}

\correspondingauthor{Abril Sahade}
\email{asahade@unc.edu.ar}

\author{Abril Sahade}
\affiliation{Instituto de Astronom\'{\i}a Te\'orica y Experimental, CONICET-UNC, C\'ordoba, Argentina.} \affiliation{Facultad de Matem\'atica, Astronom\'{\i}a, F\'{\i}sica y Computaci\'on, Universidad Nacional de C\'ordoba (UNC), C\'ordoba, Argentina.}

\author{Mariana C\'ecere}
\affiliation{Instituto de Astronom\'{\i}a Te\'orica y Experimental, CONICET-UNC, C\'ordoba, Argentina.}
\affiliation{Observatorio Astron\'omico de C\'ordoba, UNC, C\'ordoba, Argentina.}

\author{Gustavo Krause}
\affiliation{Instituto de Estudios Avanzados en Ingenier\'{\i}a y Tecnolog\'{\i}a (IDIT), CONICET, C\'ordoba, Argentina.}
\affiliation{Facultad de Ciencias Exactas, F\'{\i}sicas y Naturales, UNC, C\'ordoba, Argentina.}




\begin{abstract}

The understanding of the causes that produce the deflection of coronal mass ejections (CMEs) is essential for the space weather forecast. In this article, we study the effects on CMEs trajectories produced by the different properties of a coronal hole close to the ejection area. For this analysis, we perform numerical simulations of the ideal magnetohydrodynamics equations that emulate the early rising of the CME in presence of a coronal hole. We find that, the stronger the magnetic field and the wider the coronal hole area, the larger the CME deflection. This effect is reduced when the coronal hole moves away from the ejections region. To characterize this behavior, we propose a dimensionless parameter
that depends on the coronal hole properties and properly quantifies the deflection. Also, we show that the presence of the coronal hole near a CME magnetic structure produces a minimum magnetic energy region which is responsible for the deflection. Thus, we find a relationship between the coronal hole properties,  the location of this region and the CME deflection.

\end{abstract}



\section{Introduction}\label{s:intro} 
 Coronal mass ejections (CMEs) are eruptive events in which large amounts of solar mass are released towards the interplanetary medium. They often interact with the Earth's magnetosphere and cause geomagnetic storms, making them objects of great interest for space weather forecasting studies. It is known that CMEs do not always evolve in a radial direction but can deviate due to multiple factors dificulting the prediction of Earth encounters \citep{2017ApJ...845..117Z}.
 In this context, the analysis of the coronal environment 
 during the early stages of the CME's evolution
 is of utmost importance to estimate a probable trajectory. 
 The CMEs deviations from the radial direction (deflection, hereafter) are mainly attributed to the distribution of several magnetic structures surrounding the CME formation area, namely: coronal holes \citep[e.g.,][]{2006AdSpR..38..461C,2009SoPh..259..143X,2009JGRA..114.0A22G,2009AnGeo..27.4491K,2013SoPh..287..391P}, active regions \citep[e.g.,][]{2015ApJ...805..168K,2015NatCo...6E7135M}, pseudostreamers \citep[e.g.,][]{2013ApJ...764...87L}, streamer belts \citep[e.g.,][]{2012ApJ...744...66Z,2013ApJ...775....5K,2018ApJ...862...86Y}, and heliospheric current sheets \citep[e.g.,][]{2015SoPh..290.3343L,2019arXiv190906410W}. 

 Currently in literature, mainly coronal magnetic structures like pseudostreamers (PSs), streamer belts (SBs) or heliospheric current sheets (HCSs) are related with low magnetic energy regions. These structures act as potential wells where the CMEs are believed to ``fall'' changing their radial trajectory. On the other hand, active regions (ARs) act in a different way, since the strong gradients of the magnetic field strength present in the neighborhood of these regions are responsible for the CME deflection.
Moreover, coronal holes (CHs) seem to act as magnetic walls, because CMEs cannot penetrate their open magnetic field and are pushed in the opposite direction. 

There are several case studies \citep[see, e.g.,][]{2017ApJ...834..172J,2018ApJ...862...86Y,2019AdSpR..XX.XXXXC}, where the deflection of the analyzed CME seems to be produced by the interaction with a CH. In these studies, it is observed that the CMEs move away from the CH region. 
In studies of observational multievents, the influence of coronal holes in the CMEs trajectory is studied by means of a quantity that represents a ``fictitious force'' which depends on both, the area of the CH and its  distance to the CME \citep{2006AdSpR..38..461C}.
These studies show a good agreement between the measured deviation and the sum of all fictitious forces of the holes. 

In addition to these studies, \citet{2009JGRA..114.0A22G} analyzed the influence of coronal holes in the propagation of six CMEs using an ``influence parameter'', which includes the CH magnetic field strength. They conclude that the open field lines of the coronal holes act as a ``magnetic wall that constrains the CME propagation.''
Also, the study of a large amount of magnetic clouds (MCs) and non-MCs, show that the deflection of interplanetary CMEs is influenced by CHs \citep{2013SoPh..284...59M}. 

Many efforts have been carried out to understand the behavior of coronal mass ejections, in which the numerical simulations of the magnetohydrodynamics (MHD) equations play a preponderant role. Several
works in the literature are devoted to study the
CME deflection analyzing the ability of different magnetic structures surrounding a CME event to produce a deflection \citep[see, e.g.,][]{2011ApJ...738..127L,2012ApJ...744...66Z,2013ApJ...764...87L,2014JGRA..119.9321Z}. 
Other simulations of multicases \citep[see, e.g.,][]{2013JGRA..118.6007Z} study 
magnetic structures containing regions of low magnetic energy, imbalance in the magnetic pressure and tension,
magnetic tension and pressure gradient or reconnections that produce magnetic forces  giving rise to the CME deviation. 

A method to analyze how the trajectory of CMEs is altered is the ForeCAT model \citep{2013ApJ...775....5K,2015ApJ...805..168K}. This model takes into account the properties of the CME (e.g., mass, expansion, velocity, etc.) and analyze how the magnetic forces, produced by different background magnetic structures, affect the radial trajectory of a CME. Using this model, \citet{2017ApJ...839...37C} analyzed the deviation of the CME event of April 9, 2008, and suggested for this particular case that the prominence dynamics itself is the cause of the CME deviation.

In most observational cases the CMEs deflections have been detected at coronagraph altitudes ($>$1.5\,R$_\sun$), with a few cases at altitudes larger than $1.12\,$R$_{\sun}$ \citep[see, e.g.,][]{2010ApJ...711...75L,2011JASTP..73.1129P,2012ApJ...744...66Z,2013SoPh..287..391P,2017ApJ...839...37C}.
To study the cases where a deflection occurs, it is necessary to characterize the coronal
environment during the first evolution stages of the ejection at low coronal levels (until 1.5\,R$_{\sun}$). 
Although it is well known that the magnetic structures surrounding the CME affect its trajectory, due to the complexity of the configurations, it
is difficult to quantify the specific action of each structure of the magnetic environment on the CME trajectory.

In the present work we numerically study the effect of coronal holes in an isolated CME flux rope with the aim of characterizing the deflection as a function of the CH features and the magnetic environment configuration. For this purpose, we carry out numerical simulations of the  ideal MHD equations  in 2.5 dimensions to model the CME evolution in different magnetic scenarios with the presence of a coronal hole. This allows us to analyze the effect of each property of the CH (area, distance, and magnetic field strength) in the early stages of the CME evolution.
We think that this systematic study contributes to the understanding of the CME behavior and its interaction with coronal holes.

 \section{The Model}\label{s:modelo}

\begin{figure*}    
   \centerline{\includegraphics[width=0.75\textwidth,clip=]{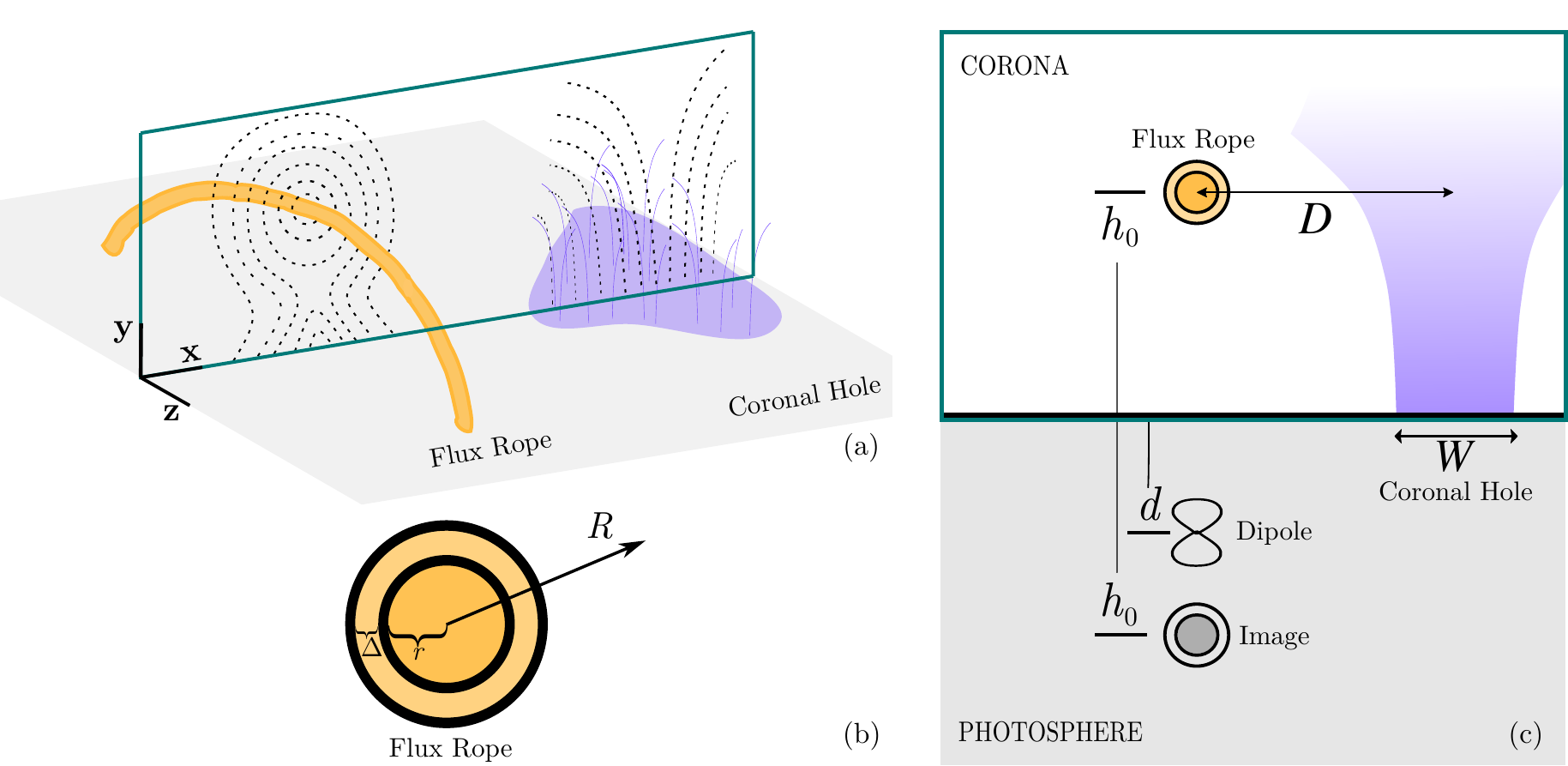}
              }
 \caption{(a) Flux rope and coronal hole scheme. Black dashed-lines represent the magnetic field, the orange thick line represents a FR, and the violet region represents a CH. Shaded plane indicates the photospheric surface. (b) Current wire layers: $r$ is the current wire radius, $\Delta$ is the thickness of the transition  layer  between  the  current wire  and  the  exterior and $R$ is the radial coordinate from the center of the current wire (in this case, the FR). (c) Components of the magnetic field with their relative two-dimensional positions.
}\label{F-corte}
\end{figure*}

The basic model starts with the ideal MHD equations in presence of a gravitational field, which arise from considering the macroscopic behavior of a compressible ideal fully ionized plasma. The ideal MHD equations for a Cartesian system in its conservative form and in CGS units are written as:
\begin{equation}\label{e:cont}
{\textstyle \frac{\partial\rho}{\partial t}+\boldsymbol{\nabla\cdot}(\rho\boldsymbol{v})=0} \, ,
\end{equation}

\begin{equation}\label{e:euler}
\!{\textstyle \!\frac{\partial (\rho \boldsymbol{v})}{\partial t} + \boldsymbol{\nabla \cdot} \left(\rho \boldsymbol{v} \boldsymbol{v} - \frac{1}{4\pi} \boldsymbol{B}\boldsymbol{B} \right) + \boldsymbol{\nabla} p + \frac{1}{8\pi} \boldsymbol{\nabla} B^2   = \rho \boldsymbol{g}} \, ,
\end{equation}
\begin{equation}\label{e:consE}
 \!\!{\textstyle \frac{\partial E}{\partial t} + \boldsymbol{\nabla \cdot} \left[\left(E + p + \frac{B^2}{8\pi}\right)\boldsymbol{v} -\frac{1}{4\pi} \left(\boldsymbol{v\cdot B}\right)\boldsymbol{B}\right] = \rho \boldsymbol{gv}} \, ,
\end{equation}
 \begin{equation}\label{e:induccion}
{\textstyle \frac{\partial \boldsymbol{B}}{\partial t} + \boldsymbol{\nabla \cdot} \left(\boldsymbol{v} \boldsymbol{B} - \boldsymbol{B} \boldsymbol{v} \right) = \boldsymbol{0}} \, ,
\end{equation}

\noindent
where $\rho$ indicates the plasma density, $p$ the thermal pressure, $\boldsymbol{v}$ the velocity, $\boldsymbol{B}$ the magnetic field, $\boldsymbol{g}$ is the gravity acceleration, and $E$ is the total energy (per unit volume) given by  
\begin{equation*}
   {\textstyle E = \rho \epsilon + \frac{1}{2} \rho v^2 + \frac{B^2}{8\pi},}
\end{equation*}
where $\epsilon$ is the internal energy and
\begin{equation*}
{\textstyle \boldsymbol{j}=\frac{c}{4\pi}\boldsymbol{\nabla\times}\boldsymbol{B}, } \,
\end{equation*}
is the current density, with $c$ the speed of light.

In addition to the MHD equations, the divergence-free condition of the magnetic field, i.e.
\begin{equation}\label{e:divB}
 {\textstyle \boldsymbol{\nabla\cdot} \boldsymbol{B} = 0}\, ,
\end{equation}
must be fulfilled.

To complete the set of MHD equations a closure relation among the thermodynamic variables must be imposed. We assume a calorically perfect gas for which $p = 2\rho k_B T/m_i = (\gamma - 1) \rho \epsilon$, where $k_B$ is the Boltzmann constant, $T$ the plasma temperature, $m_i$ the proton mass (assuming that the plasma is fully ionized hydrogen), and $\gamma = 5/3$ is the specific heats relation. 

\subsection{Numerical code}
In order to evaluate the plasma behavior, the MHD equations~\eqref{e:cont}--\eqref{e:induccion} are numerically solved in a two-dimensional Cartesian grid of co-located finite volumes.
We perform 2.5D simulations to consider the $z$-direction magnetic field component in the interior of the flux rope, which are carried out using the FLASH Code \citep{2000ApJS..131..273F}, an open-source publicly available suite of high-performance simulation tools developed at the Center for Astrophysical Thermonuclear Flashes (Flash Center) of the University of Chicago.
This code, currently in its fourth version, uses the finite volume method with Godunov-type schemes to solve the high energy compressible MHD equations on regular grids with adaptive mesh refinement (AMR) capabilities. For our simulations we choose the USM (unsplit staggered mesh) solver available in FLASH, for which it uses a second-order directionally unsplit scheme with a MUSCL-type reconstruction. This solver implements a more consistent treatment of the magnetic field, since its formulation is based on the constrained transport method and the corner transport upwind method, which avoids the generation of non-physical magnetic field divergence \citep{2009JCoPh.228..952L}.
To solve the interface Riemann problems we set the Roe's solver among the available options.

Cartesian 2D rectangular grids are used to represent the physical domain  of $[-1\,000,1\,000]\text{Mm}\times [0,1\,000]\text{Mm}$ with an initial $80 \times 40$ discretization and nine refinement levels that take into account the pressure and temperature gradients.
With this discretization we obtained a resolution of $\sim [0.1 \times 0.1]\text{Mm}^2$ for the maximum refinement.
Boundary conditions are set as follows. At both lateral ends outflow conditions (zero-gradient) are applied for the thermodynamic variables and the velocity to  allow waves to leave the domain without reflection. The boundary conditions of the magnetic field at lateral ends require to extrapolate the initial force-free configuration to ghost cells in order to avoid the generation of spurious magnetic forces produced when assuming a zero-gradient evolution extrapolation in a non-constant magnetic field. Obviously, this model is valid as long as shocks or disturbances do not reach the lateral ends of the domain. In the lower and the upper limits the hydrostatic boundary conditions must be imposed due to the effect of gravity, which acts in the $y$-direction and produces spurious fluxes through the top and the bottom of the domain if pressure and density are not correctly extrapolated. Therefore, to guarantee the conservation of the hydrostatic equilibrium at both ends we use the extrapolation proposed by \citet{2019A&A...631A..68K} considering constant temperature through the boundary.
The remaining variables (velocity and magnetic field components) at the upper boundary are extrapolated with a zero-gradient assumption, while for the lower boundary we impose the condition described by \citet{1987SoPh..114..311R} to ensure the line-tied magnetic field condition that is present in the solar surface during the CME evolution.

In addition to the described features of the numerical model, a particular treatment is required to correctly simulate the strong stratification of the background atmosphere. Then, to avoid spurious vertical velocities associated to the unbalance between the numerical fluxes and the discrete gravitational source term when standard MUSCL-type schemes are used in strongly stratified atmospheres, we implement the local hydrostatic reconstruction scheme proposed by \citet{2019A&A...631A..68K} to improve the preservation of hydrostatic equilibrium during the simulation.

Regarding the diffusive effects, we already established that the ideal MHD equations are used, therefore there is no physical diffusion added to our experiments.
Analytically, the ideal MHD model do not permit the magnetic reconnection, thus a current sheet should be formed in the region below the flux rope where the magnetic lines are strongly stretched, which eventually causes the stopping of the flux rope rising \citep{1990JGR....9511919F}. However, the numerical diffusion present in the simulations provides the necessary disipation to prevent the current sheet formation allowing the ejection.
We performed this analysis in a previous paper \citep{2018MNRAS.474..770K} where it is showed that the presence of anomalous magnetic resistivity in the region of the current sheet formation do not change the ejection velocity of the flux rope with respect to the ideal model. In this way, we can neglect the magnetic resistivity and use the ideal MHD equations, which allows a significant reduction in the computational cost.

\subsection{Stratified atmosphere}

To simulate the solar atmosphere we adopt a multi-layer atmosphere structure \citep{2012MNRAS.425.2824M}. The chromosphere is located between $y=0$ and $y=h_{\text{ch}}$ with constant temperature $T_{\text{ch}}$. Above this, the transition region is extended to the base of the corona ($y=h_{\text{c}}$) where the temperature grows linearly until $T_{\text{c}}$, the constant temperature of the corona. Then the initial temperature distribution is given by
\begin{equation}
    {\textstyle T{\scriptstyle(y)}=}
    \left\{
\begin{array}{rl}
\begin{alignedat}{2}
&{\textstyle T_{\text{ch}}} &&\text{\small if } {\scriptstyle 0\leq y < h_{\text{ch}}}\\
&{\textstyle (T_{\text{c}}-T_{\text{ch}})\left[\frac{y-h_{\text{ch}}}{h_{\text{c}}-h_{\text{ch}}}\right]+T_{\text{ch}}} \quad && \text{\small if } {\scriptstyle h_{\text{ch}}\leq y < h_{\text{c}}}\\
& {\textstyle T_{\text{c}}}  && \text{\small if } {\scriptstyle h_{\text{c}} \leq y }.
\end{alignedat}
\end{array} \right.  
\end{equation}

We set up a temperature of $T_{\text{ch}}=10\,000\,\text{K}$ at the chromosphere and $T_{\text{c}}=10^6\,\text{K}$ at the corona. The height of the chromosphere is  $h_{\text{ch}}=10\,\text{Mm}$, the transition region extends for $5\,\text{Mm}$ until $h_{\text{c}}=15\,\text{Mm}$, the base of the corona.

Considering the atmosphere in hydrostatic equilibrium and current free, the pressure is obtained from the combination of the equation of state and eq.~\eqref{e:euler} with $\boldsymbol{v} = \boldsymbol{0}$. Then, taking a system with the $y$-axis aligned to the gravity acceleration but in the opposite direction (i.e., $\boldsymbol{g} = (0, -G M_\sun/(y + R_\sun)^2, 0)$, where $G$ is the gravitational constant, $M_\sun$ is the Sun's mass, $R_\sun$ is the solar radius, and $y = 0$ corresponds to the solar surface), the hydrostatic pressure distribution is only a function of $y$: 
\begin{equation*}
    {\textstyle  p{\scriptstyle(y)}=\!}
    \left\{
\begin{array}{rl}
\begin{alignedat}{2}
&{\textstyle \!\!p_{\text{ch}}\exp{\! \left[\left(\frac{h_{\text{ch}}}{1+h_{\text{ch}}/R_{\sun}}-\frac{y}{1+y/R_{\sun}}\right)\frac{\alpha}{T_{\text{ch}}}\right]}} &&\text{\small if } {\scriptstyle 0\leq y < h_{\text{ch}}}\\
&{\textstyle\! \!p_{\text{ch}}\exp{\!\left[-\int_{h_{\text{ch}}}^{y}\frac{\alpha}{T{\scriptstyle(y')}}\left(1+\frac{y'}{R_{\sun}}\right)^{-2}\!\! dy'\right]}} \quad && \text{\small if } {\scriptstyle h_{\text{ch}}\leq y < h_\text{c}}\\
&{\textstyle \! \!p_{\text{c}}\exp{\!\left[-\frac{(y-h_{\text{c}})}{1+(y-h_{\text{c}})/R_{\sun}}\frac{\alpha}{T_{\text{c}}}\right]}}  && \text{\small if } {\scriptstyle  h_{\text{c}}\leq y}

\end{alignedat}
\end{array} \right.  
\end{equation*}
where 
\begin{equation*}
    {\textstyle p_{\text{ch}}{\scriptstyle(y)}=p_{\text{c}}\exp{\left[\int_{h_{\text{ch}}}^{h_{\text{c}}}\frac{\alpha}{T(y')}\left(1+\frac{y'}{R_{\sun}}\right)^{-2} dy'\right]}} \, ,
\end{equation*}
and $\alpha = m_ig_{\sun}/2k_B$, with $g_\sun = G M_\sun/R_\sun^2$.

The associated density is obtained from the equation of state, i.e.:
\begin{equation}
{\textstyle \rho=\frac{m_ip{\scriptstyle(y)}}{2k_BT{\scriptstyle(y)}}}.    
\end{equation}

\subsection{CME model}

The catastrophe model by \citet{1990JGR....9511919F} consists of a magnetic configuration out of equilibrium driving the ejection of the flux rope (FR hereafter). Forbes proposed that the magnetic field of the FR is produced by a current wire (originally proposed by \cite{1978SoPh...59..115V}). An image current wire is located below the photosphere with opposite direction to generate a repulsive force. Also, the model includes a line dipole below the photosphere which provides an attractive force to the CME's wire and emulates the photospheric field. Figure~\ref{F-corte}(a) shows a scheme of the magnetic field (black dashed-lines) of a flux rope (orange thick line) and a coronal hole (violet region). In Figure~\ref{F-corte}(b) a current wire (the FR or the image current wire) is schematized, where we can describe three zones \citep{2012MNRAS.425.2824M}:
\begin{description}
    \item[Z1] Inside a current wire, $0\leq R < r-\frac{\Delta}{2}$.
    \item [Z2] Throughout the transition layer, $r-\frac{\Delta}{2}\!\leq \!R\! <r+\frac{\Delta}{2}$.
    \item [Z3] Outside a current wire, $r+\frac{\Delta}{2} \leq R$,
\end{description}
where $r$ is the current wire radius and $\Delta$ is the thickness of the transition layer between the current wire and the exterior and $R$ the radial coordinate from the center of the current wire.

The magnetic field component $B_\phi$ generated by a current wire with current distribution $j_z$, which are given by eq.~\eqref{e:Bphi} and \eqref{e:jz}:

\begin{equation} \label{e:Bphi}
   \!\!\!\!\!\! B_\phi{\scriptstyle(R)}\!=\!
    \left\{
\begin{array}{rl}
\begin{alignedat}{2}
&\tfrac{2\pi}{c}j_0R  &&\text{\small at {\bf Z1}}\\
&\tfrac{2\pi j_0}{cR}\left\{\tfrac{1}{2}\left(r-\tfrac{\Delta}{2}\right)^2-\left(\tfrac{\Delta}{2}\right)^2+\right. 
\\
&\tfrac{R^2}{2}+\tfrac{\Delta R}{\pi}\text{sin}\left[\tfrac{\pi}{\Delta}\left(R-r+\tfrac{\Delta}{2}\right)\right]+

\\
&\left.\!\!\!\left(\tfrac{\Delta }{\pi}\right)^2\cos\left[\tfrac{\pi}{\Delta}\left(R-r+\tfrac{\Delta}{2}\right)\right]\right\} \qquad && \text{\small at {\bf Z2}}\\
&\tfrac{2\pi j_0}{cR}\left[r^2+\left(\tfrac{\Delta}{2}\right)^2-2\left(\tfrac{\Delta}{\pi}\right)^2\right]  && \text{\small at {\bf Z3},}
\end{alignedat}
\end{array} \right.
\end{equation}
\begin{equation} \label{e:jz}
   \! j_z{\scriptstyle(R)}\!=\!
    \left\{
\begin{array}{rl}
\begin{alignedat}{2}
&\!j_0 &&\text{\small at {\bf Z1}}\\
&\!\tfrac{j_0}{2}\left\{\cos\left[\tfrac{\pi}{\Delta}\left(R-r+\tfrac{\Delta}{2}\right)\right]+1\right\} \quad  && \text{\small at {\bf Z2}}\\
&\! 0  && \text{\small at {\bf Z3};}
\end{alignedat}
\end{array} \right.
\end{equation}
\noindent
where $j_0$ is a current density. 

In order to obtain a helical magnetic field in the FR, to achieve densities and temperatures consistent with observational data of flux ropes, we include to the catastrophe model a magnetic field in $z$-axis of strength $B_z$. In this way, we avoid excessive gas pressure values needed to balance the magnetic pressure inside the flux rope in the initial equilibrium state. The component $B_{\text{z}}$ of the magnetic field, and the current distribution $j_{\phi}$, are described by:
\begin{equation}\label{e:Bfield}
    B_{\text{z}}{\scriptstyle(R)} = \tfrac{\sqrt{8}\pi j_1}{c}\sqrt{\left(r-\tfrac{\Delta}{2}\right)^2-R^2}\, , 
\end{equation}
\begin{equation}\label{e:jphi}
    j_\phi{\scriptstyle(R)} = j_1R\left[\sqrt{\left(r-\tfrac{\Delta}{2}\right)^2-R^2}\right]^{-1}\,, 
\end{equation}
where $j_1$ is a current density. These expressions are valid inside the flux rope ({\bf Z1}) and are null in the rest of the domain. 

Then, the Cartesian components of the magnetic field in the whole computational domain are given by \citep{2012MNRAS.425.2824M}:
\begin{align}\label{e:BfieldFR}
    B_x=&-B_\phi{\scriptstyle(R_-)}\tfrac{(y-h_0)}{R_-} +B_\phi{\scriptstyle(R_+)}\tfrac{(y+h_0)}{R_+} -  \nonumber \\
    & \: MdB_\phi{\scriptstyle\left(r+\tfrac{\Delta}{2}\right)}\left(r+\tfrac{\Delta}{2}\right)\tfrac{x^2-(y+d)^2}{R_d^4} \, ,\nonumber \\
    B_y= &B_\phi{\scriptstyle(R_-)}\tfrac{x}{R_-} -B_\phi{\scriptstyle(R_+)}\tfrac{x}{R_+}-  \nonumber \\
    & MdB_\phi{\scriptstyle\left(r+\tfrac{\Delta}{2}\right)}\left(r+\tfrac{\Delta}{2}\right)\tfrac{2x(y+d)}{R_d^4}  \, , \nonumber \\
    B_z=&B_{\text{z}}{\scriptstyle(R_-)}\, .
\end{align}
\noindent
where $h_0$ is the initial vertical position of the FR and $M$ is the intensity of the line dipole at depth $d$.
 The distances $R$ are:
\begin{equation*}
    R_\pm = \sqrt{x^2+(y\pm h_0)^2},
\end{equation*}
\begin{equation*}
    R_d = \sqrt{x^2+(y+d)^2},
\end{equation*}
where $R_{-}, R_{+}$ and $R_{d}$ are taken from having their origins in the FR, image current wire and line dipole, respectively. In this way, the first, second and third terms, for example of $B_x$ correspond to the $x$-component of the magnetic field produced by the FR, image current wire and line dipole, respectively. Fig.~\ref{F-corte} (c) shows a scheme with the relative positions between the components of the magnetic field, including the coronal hole. 

The temperature inside the FR ($T_{\text{\tiny{FR}}}$) varies according to the following temperature distribution:
\begin{equation}
    \!T{\scriptstyle(R_-)}\!=\!
    \left\{
\begin{array}{rl}
\begin{alignedat}{2}
&\!T_{\text{\tiny{FR}}} && \text{\small at {\bf Z1}}\\
&\!\!(T_{\text{c}}\!-\!T_{\text{\tiny{FR}}})\!\left[\tfrac{R_--(r+\Delta/2)}{\Delta}\right]\!+\!T_{\text{\tiny{FR}}} \quad && \text{\small at {\bf Z2}}\\
&\! T_{\text{c}} && \text{\small at {\bf Z3}.}
\end{alignedat}
\end{array} \right.
\end{equation}
\noindent

The internal pressure of the FR is obtained by proposing a solution close to equilibrium:
\begin{align} \label{e:presion}
    p_{\text{\tiny{FR}}}{\scriptstyle(x,y)} = & p{\scriptstyle(y)}+\tfrac{1}{c}\int_{R}^{r+\frac{\Delta}{2}}B_\phi{\scriptstyle(R')}j_z{\scriptstyle(R')}dR'\nonumber\\
    &-\tfrac{1}{c}\int_{R}^{r+\frac{\Delta}{2}}B_{\text{z}}{\scriptstyle(R')}j_\phi{\scriptstyle(R')}dR',
\end{align}
where $p{\scriptstyle(y)}$ is the background hydrostatic pressure.

We consider that the flux rope length is large enough in order to satisfy the 2.5D assumption, which is in agreement with the observations where lengths of $(100-500)\,$Mm  are registered \citep{2014IAUS..300...15B}.

 \subsection{Coronal hole model}\label{sss:chmodel}
 
\begin{figure}    
    \includegraphics[width=0.4\textwidth,clip=]{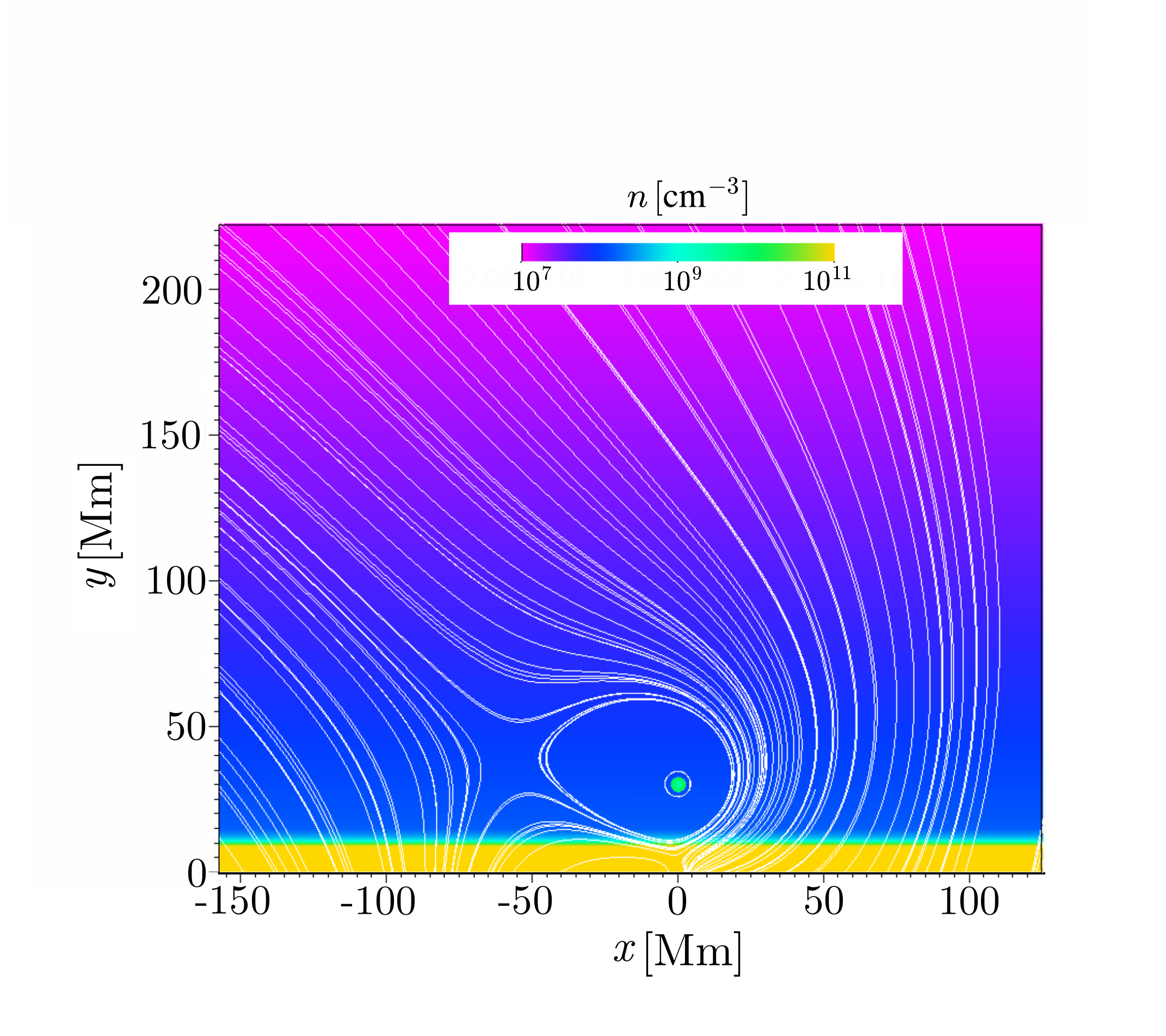}
    \caption{Initial density distribution (color-map) and magnetic field lines (white lines). An animated version of this figure, showing the magnetic field lines and density evolution, is available in the HTML version.}
                      
   \label{F-magfield}
   
   \end{figure}

The initial magnetic field of the coronal hole is described by \citep{2014A&A...568A..20P}:
\begin{align}\label{e:BfieldCH}
    B_x= &B_0\ \sin\left(\frac{x-D}{W}\right)\, e^{-y/W} \, ,\nonumber \\
    B_y= &B_0\ \cos\left(\frac{x-D}{W}\right)\,e^{-y/W} \, ,\nonumber \\
    B_z= &0.
\end{align}

The parameter $B_0$ is the radial magnetic field strength of the coronal hole at the distance $D$ on the $x$-axis. The parameter $W$ is related to the width of the coronal hole, and modifies the decay of the strength of magnetic field in the $y$-direction.
 
 The total initial magnetic field is the sum of the magnetic field of the FR (eqs.~\eqref{e:BfieldFR}) and the corresponding to the CH (eqs.~\eqref{e:BfieldCH}). Fig.~\ref{F-magfield} shows the total magnetic field in white lines and the plasma density distribution in color-map. As can be noted, a point of minimum magnetic energy located at the left of the flux rope ($\sim(-50,50)$Mm) is present in this configuration. The FR evolves toward the minimum magnetic energy region, as can be seen in the animated version of Fig.~\ref{F-magfield}.  The relation between the location of the minimum magnetic energy (MME) region and the CME deflection is analyzed with more detail in Section~\ref{Results}.
 
  \subsection{Parameter selection}\label{ss:parametros}
To this study, we simulate warm flux ropes (with temperatures equal to or greater than coronal ones), outside active regions, at a height of $30\,$Mm with a diameter of $5\,$Mm \citep{2014IAUS..300...15B}. These FRs usually have extensions of $(100-500)\,$Mm, therefore we consider a characteristic length of $L_0=100\,$Mm in the \textit{z}-direction. By the proposed model we obtain number density values inside the FR in the range of $(5\times 10^8-1\times 10^{10})\,\text{cm}^{-3}$,  which are comparable to the observations reported by \cite{2012ApJ...761...62C,2016A&A...588A..16S}. To achieve $B\sim 1 \,\text{G}$ nearby, for the background magnetic field, we choose the relative intensity and depth of the dipole as were used in \citet{2018MNRAS.474..770K}.

Our purpose is to analyze how the different characteristics of a coronal hole influence the deflection of coronal mass ejections. To do this, we change the parameters that define the CH and analyze how these changes modify the rising trajectory of the CME. In addition, we want to analyze the dependence of the FR's configuration with regard to the modifications of the CHs. For this, we study the influence of the CH in two different flux ropes (FR1, FR2) embedded in two different coronal environments, whose number density values at the base of the corona are: $n_{\text{c,\tiny{FR1}}}=3 \times 10^8\,\text{cm}^{-3}$ and $n_{\text{c,\tiny{FR2}}}=4.5 \times 10^8\,\text{cm}^{-3}$ \citep{2010ApJ...725.1373V,2016AdSpR..57.1286V}. The two simulated FRs have different current densities $j_0$ and $j_1$, whose magnetic field values are between $(10-100)\,\text{G}$ inside them. 

As described in Section~\ref{sss:chmodel}, the parameters of a CH are $B_0$, $D$, and $W$. We choose values to obtain typical non-polar CH scenarios. Statistical studies found the absolute value of the magnetic field strength to be distributed from $0.2\,$G to $14.0\,$G, with areas ($A\sim W L_0$) between ($1.6\times 10^3 - 1.8\times 10^{5})\,\text{Mm}^2$ \citep{2017ApJ...835..268H,2019SoPh..294..144H}. 

Under these considerations, we perform a parametric study varying the CH features, which is carried out considering the values of Table~\ref{tbl-1}. In Table \ref{tbl-2} we show the two configurations of the FR1 and FR2, and the remaining fixed parameters. 
\begin{deluxetable}{c  c c c}
\tablecaption{Coronal holes parameters for each case.\label{tbl-1}}
\tablewidth{0pt}
\tablehead{
\colhead{Case} & \colhead{$B_0\,$[G]} &  \colhead{$D\,$[Mm]} & \colhead{$W\,$[Mm]} 
}

\startdata
    1 & $0.4$  & $150$ & $400$\\
    2 & $0.8$  & $150$ & $400$ \\
    3 & $1.2$  & $150$ & $400$\\
    4 & $1.6$  & $150$ & $400$\\
    5 & $0.8$  & $180$ & $400$\\
    6 & $0.8$  & $250$ & $400$\\
    7 & $0.8$  & $350$ & $400$\\
    8 & $0.8$  & $150$ & $300$\\
    9 & $0.8$  & $150$ & $500$ \\
    10 & $0.8$ & $150$ & $600$ 
\enddata
\end{deluxetable}

\begin{deluxetable}{c c c}
\tablecaption{Initial state parameters.\label{tbl-2}}

\tablehead{
\colhead{Parameter}  & \multicolumn{2}{c}{Value}\\
\nocolhead{}& \colhead{FR 1}& \colhead{FR 2}
}

\startdata
 $j_0\,[\text{stA}\,\text{cm}^{-2}]$  & $435$& $525$ \\
  $j_1\,[\text{stA}\,\text{cm}^{-2}]$ & $455$& $300$ \\
  $T_{\text{\tiny{FR}}}\,[\text{MK}]$& $1$& $4$\\
  $n_{\text{c}}\,[\text{cm}^{-3}]$ & $3\times10^8$& $4.5\times10^8$\\
  $h_0\,[\text{Mm}]$      &\multicolumn2c{$30$}  \\ 
  $r\,[\text{Mm}]$        &\multicolumn2c{$2.5$}\\
  $\Delta\,[\text{Mm}]$        & \multicolumn2c{$0.25$} \\
  $d\,[\text{Mm}]$   & \multicolumn2c{$3.125$} \\
  $M$        & \multicolumn2c{$1$}  
\enddata
\tablecomments{Parameters $j_0$ and $j_1$ are the current densities inside the flux rope in $z$-direction and in $\phi$-direction, respectively, $T_{\text{\tiny{FR}}}$ is the internal FR temperature, $n_{\text{c}}$ is the numerical density at the base of the corona, $h_0$ is the vertical position (height) of the FR, $r$ is its radius, and $\Delta$ is the thickness of the transition layer between the FR interior and the corona. Parameters $d$ and $M$ are the depth of the line dipole below the boundary surface and its relative intensity, respectively.}
\end{deluxetable}

\begin{figure}    
   \centerline{\includegraphics[width=0.35\textwidth,clip=]{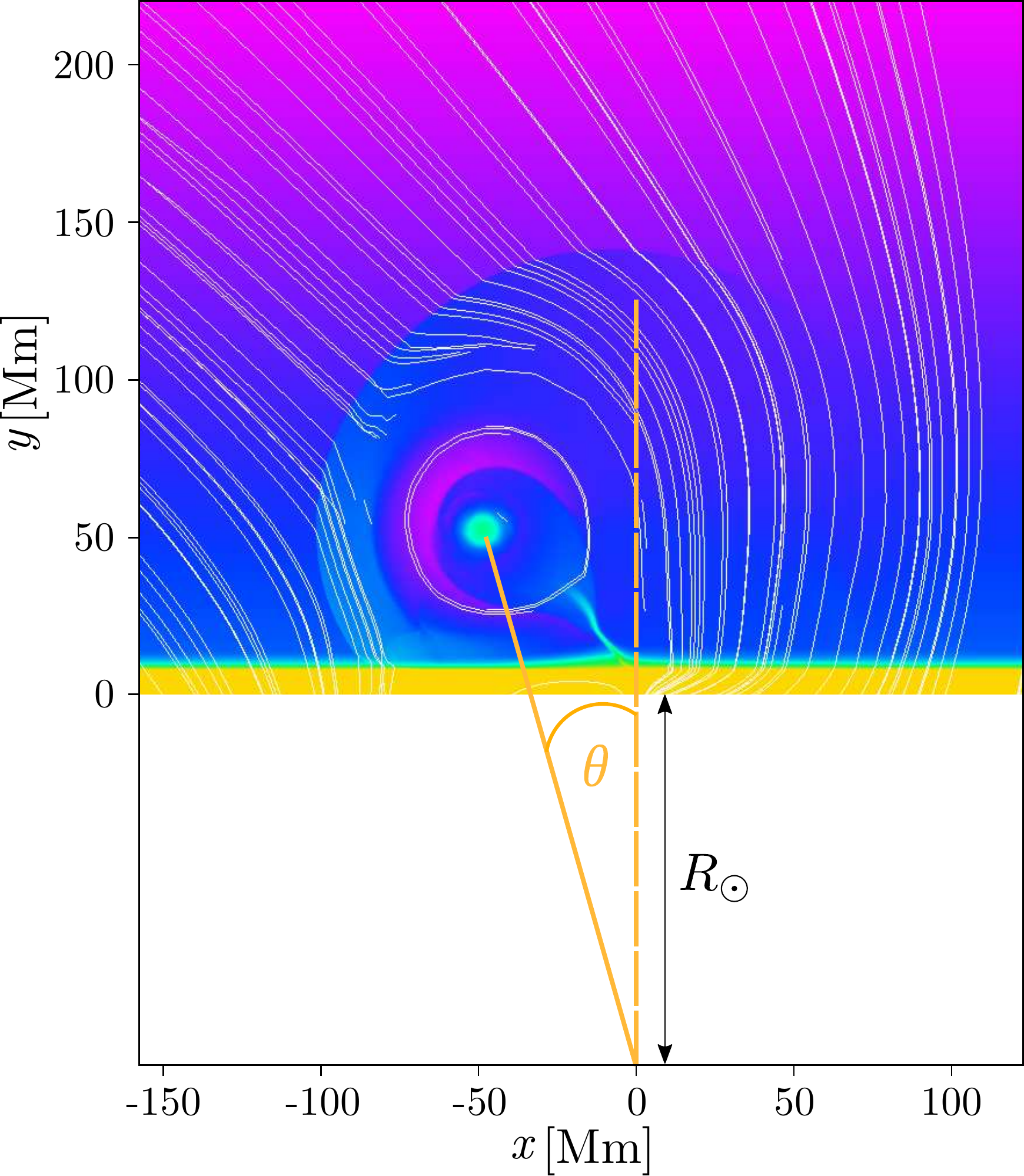}
              }
              \caption{Measurement of the angular deviation of the FR center from the radial direction (deflection angle). The $R_\sun$ is the solar radius. The color scale and white lines are like in Fig. \ref{F-magfield}.
                      }
   \label{F-deflexangle}
   \end{figure}

\section{Results}
\label{Results}
 
In this section, we present the analysis of the performed numerical simulations with the aim of quantifying the influence of the CH parameters in the deviation of the CME trajectory from the radial direction, and to understand how this deflection is driven by a coronal hole. In addition, we also analyze the relation between the location of the minimum magnetic energy region and the CME deflection.  Given a set of CH parameters, we separately analyze the dynamics of the FR1 and the FR2 configurations.

\subsection{Deflection dependence on CH parameters}
To quantify the deflection we measure the angle formed between the vertical line that passes through the initial position of the FR center, and the line defined by the position of the densest point of the FR and the solar center, as shown in Fig.~\ref{F-deflexangle}. 
Case 2 (see Table \ref{tbl-2}) is the reference coronal hole with magnetic field strength $B_{\text{{\tiny ref}}}=0.8\,$G and width $W_{\text{{\tiny ref}}}=400\,$Mm at a distance of $D_{\text{{\tiny ref}}}=150\,$Mm. We simulate the first 600 seconds (i.e. 10 minutes) of the evolution to analyze the early development of the CME. In order to compare the influence of each parameter on the deflection, we normalize the results with respect to the angle obtained by the FR in case 2 at the final time of the simulation, i.e. $\theta_0=\theta_{\text{{\tiny ref}}}(t=600\,\text{s})$. 
It has to be noted that there are two different reference deflection angles corresponding to each FR1 and FR2 scenario. With these definitions, we can evaluate the relative effect of each parameter of the CH in the CME evolution, and compare the results for each FR scenario.

\begin{figure*}    
   \centerline{\includegraphics[width=0.9\textwidth,clip=]{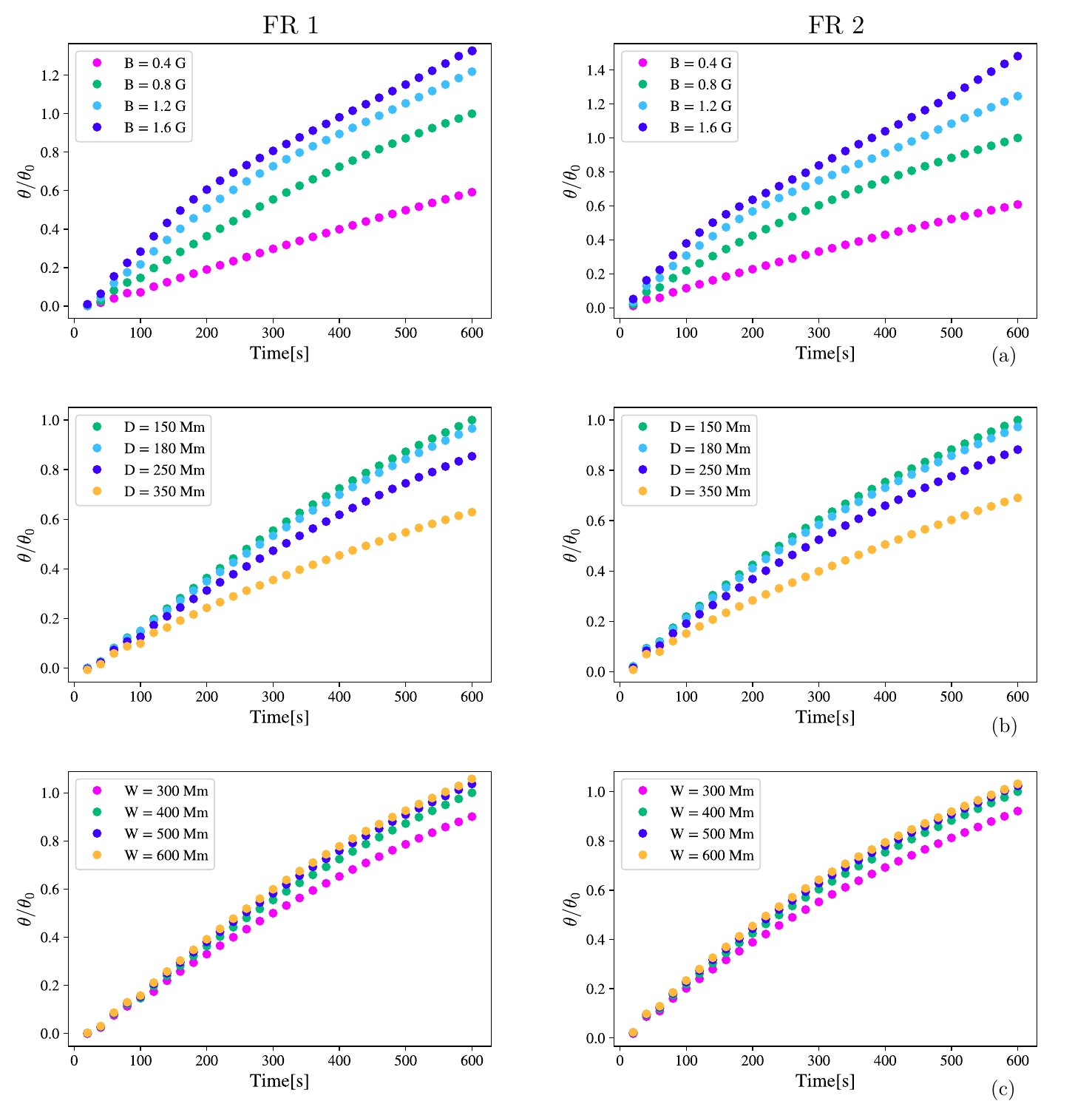}
              }
              \caption{Angular deviation of the FR's center from the vertical direction (deflection), as a function of time for different (a) magnetic field strengths, (b) distances, and (c) widths of CH. 
             }
   \label{F-deflection}
   \end{figure*}

In Fig.~\ref{F-deflection} we show the deviations obtained for FR1 (left panels) and FR2 (right panels) for the different CH configurations.
It is noticeable that the influence of the different coronal holes on both flux rope configurations is similar. All the cases showed a quasi-linear tendency, with a low deceleration. The results could be fitted by a quadratic function ($\theta/\theta_0=a t^2 + v t + c$). A more detailed description in terms of the relative deflection velocities $v/v_0$ (with $v_0$ the velocity of the reference case) is as follows:
 
\begin{itemize}
    \item The top panels of Fig.~\ref{F-deflection} show the results for cases 1, 2, 3, and 4, for which we simulate coronal holes of width $W_{\text{{\tiny ref}}}=400\,$Mm at a distance $D_{\text{{\tiny ref}}}=150\,$Mm measured from the flux rope, with different strengths of their magnetic field ($B_0=[0.4, 0.8, 1.2, 1.6]\,\text{G}$). The CH of Case 1 ($B_0=0.4\,\text{G}$) triggers a deflection $52\,\%$ and $49\,\%$ slower than the reference case (Case 2) for FR1 and FR2, respectively. For Case 3 ($B_0=1.2\,\text{G}$), the CH leads a deflection velocity $1.3$ and $1.15$ faster than the reference case for FR1 and FR2, respectively. And, for Case 4 ($B_0=1.6\,\text{G}$), the CH leads a deflection velocity $1.5$ and $1.2$ faster than the reference case for FR1 and FR2, respectively.
    \item Cases 2, 5, 6 and 7 are shown in the middle panels of Fig.~\ref{F-deflection}. These CH configurations have the same magnetic field strength ($B_{\text{{\tiny ref}}}=0.8\,\text{G}$) and the same width ($W_{\text{{\tiny ref}}}=400\,\text{Mm}$), with different distances from the FR ($D=[150, 180, 250, 350]\,\text{Mm}$). The CH of Case 5 ($D=180\,\text{Mm}$) triggers a deflection $4\,\%$ slower than the reference case (Case 2) for both FRs. In Case 6 ($D=250\,\text{Mm}$) the deflection velocity presents a reduction with respect to the reference case of $15\,\%$ for FR1 and $14\,\%$ for FR2. For Case 7 ($D=350\,\text{Mm}$) the deflection velocity decreases $36\,\%$ and $34\,\%$, respectively.
    \item The bottom panels of Fig.~\ref{F-deflection} show the results of cases 8, 2, 9 and 10, where the simulated coronal holes have the same magnetic field strength ($B_{\text{{\tiny ref}}}=0.8\,\text{G}$) and are located at the same distance from the flux rope ($D_{\text{{\tiny ref}}}=150\,$Mm), with different widths ($W=[300, 400, 500, 600]\,\text{Mm}$). The results for the coronal hole of Case 8 ($W=300\,\text{Mm}$) exhibit a reduction of the deflection velocity of $10\,\%$ for FR1 and $9\,\%$ for FR2 in relation to the reference case. For Case 9 ($W=500\,\text{Mm}$) the deflection velocity increases in a factor of $1.06$ for both FR1 and FR2. Finally, the Case 10 ($W=600\,\text{Mm}$) show an increment in deflection velocity of $1.1$ for both FRs.
\end{itemize}

\begin{figure}    
   \centerline{\includegraphics[width=0.45\textwidth,clip=]{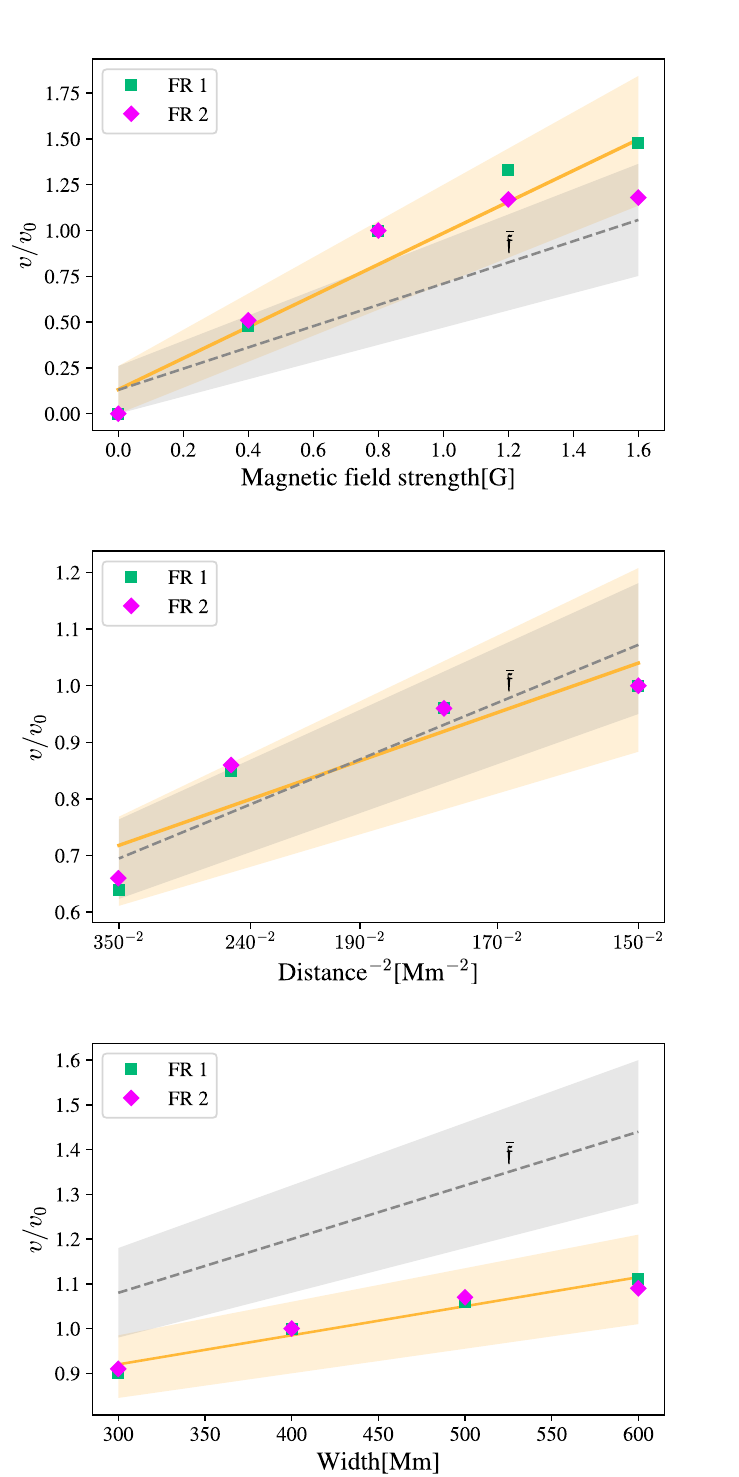}
              }
              \caption{Linear fit for the CH's parameters (orange line) with their standard deviation (orange shadow area). The black dashed-line  represents the dimensionless parameter average with its standard deviation (gray shadow area).}
   \label{F-Golpaswamy}
   \end{figure}

Following the idea of the ``influence parameter'' calculated by \citet{2009JGRA..114.0A22G} where $f \sim B_0A/D^2$, and given the quasi-linear trend of the relative deflection velocities $v/v_0$ described above, we analyze how this dimensionless quantity (named $\mathfrak{f}$ hereafter) is a function of $B_0$, $W$ and $D^{-2}$. Fig.~\ref{F-Golpaswamy} shows the linear fit (orange filled line) and its standard deviation (orange shadow area) obtained for the different CHs parameters. The top, middle, and bottom panels show the fits:
\begin{align*}\label{e:BfieldCH}
    &\mathfrak{f}_B=\alpha_B W_{\text{{\tiny ref}}} D_{\text{{\tiny ref}}}^{-2} B_0 \, , \\
    &\mathfrak{f}_D=\alpha_D W_{\text{{\tiny ref}}} B_{\text{{\tiny ref}}} D^{-2} \, , \\
    &\mathfrak{f}_W=\alpha_W B_{\text{{\tiny ref}}} D_{\text{{\tiny ref}}}^{-2} W \, ,
\end{align*}
taking into account the results of both FRs. The green square and magenta diamond symbols represent the values of $v/v_0$ obtained from FR1 and FR2, respectively. From these fits we obtained $\alpha_B = (48 \pm 8), \alpha_D = (31 \pm 7), \text{and} \  \alpha_W = (18 \pm 3)$ in units of [Mm/G]. With black dotted-line we show the average  $\bar{\mathfrak{f}}=\bar{\alpha}\frac{BW}{D^2}$, where $\bar{\alpha}=(32 \pm 4)\text{[Mm/G]}$ is the average of $\alpha_i$, with $i=B_0,D,W$. This dimensionless parameter average is in agreement with the fit values obtained for the magnetic field strength and the inverse square of the distance, but overestimates the relative deflection velocities obtained for the variation of the width ($\bar{\alpha}>\alpha_W$).

\subsection{Deflection dependence on the MME position}

\begin{figure*}    
   \centerline{\includegraphics[width=0.9\textwidth]{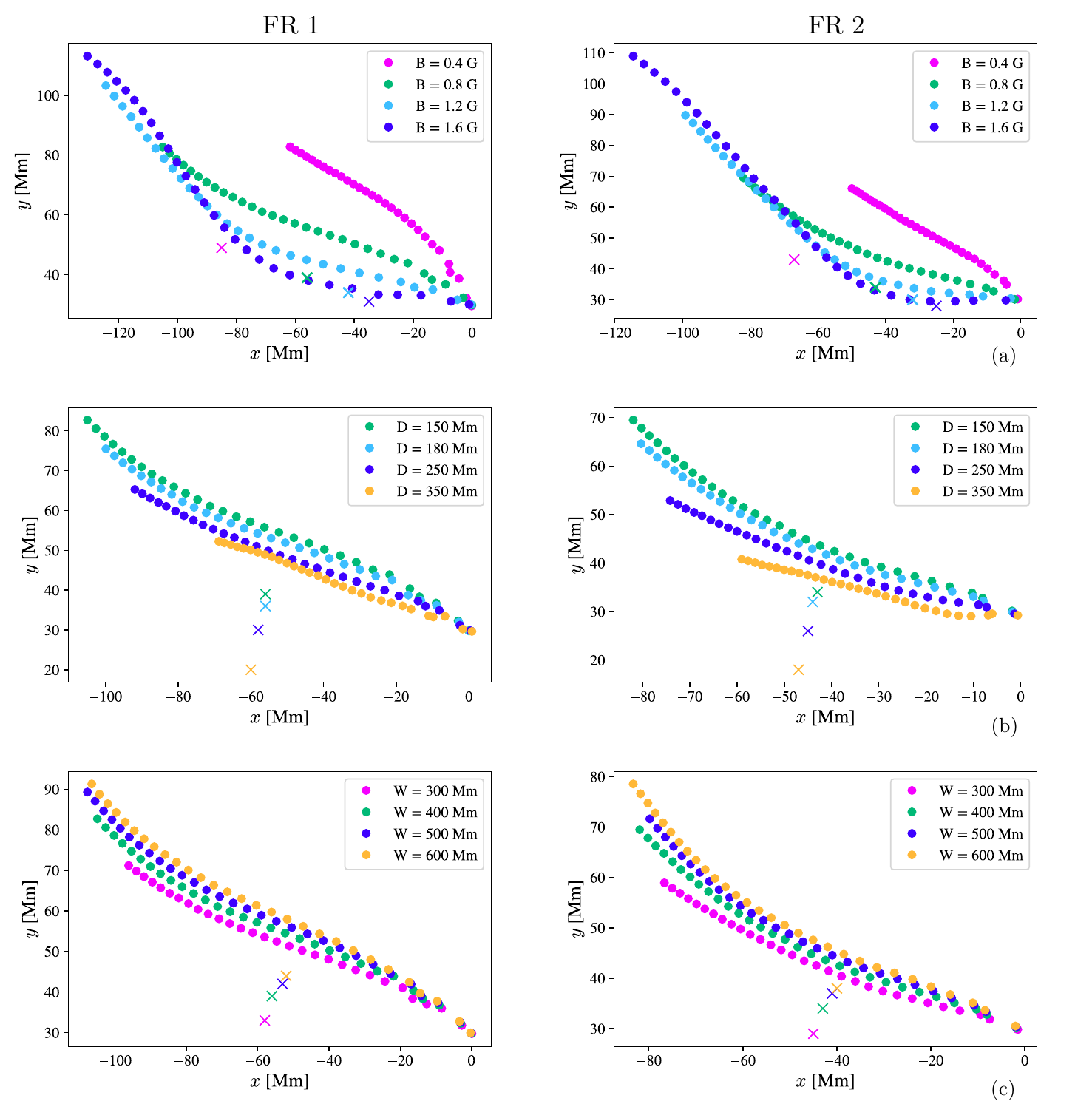}
              }
              \caption{ Path of the FR center (dots) up to $t=600\,$s, and the position of the MME at $t=0\,$s (x), for different (a) magnetic field strengths, (b) distances, and (c) widths of the CH.}
   \label{F-Energy}
\end{figure*}

 In previous works it is mentioned that coronal holes can act as magnetic walls that avoid the radial evolution of CMEs \citep{2009JGRA..114.0A22G}, but there is not enough explanation about the physical mechanisms involved in this process. As shown in Section \ref{sss:chmodel}, the presence of the coronal hole in the magnetic configuration generates a region of minimum magnetic energy to the left of the FR. To understand which is the role of this null point in the CME deflection, in Figure \ref{F-Energy} we plot the FR path for each case, up to time $t=600\,$s, overlapped to the position of the corresponding minimum of magnetic energy at the initial time (``x'' marks in the plots).

 In this analysis we can see again that even though both flux ropes follow different paths, they show a similar behavior under the variation of CH parameters. Below we analyze the path of each case according to the variation of the used parameters:
\begin{itemize}
    \item  The top panels of Fig.~\ref{F-Energy} show the paths for cases 1, 2, 3, and 4, where the magnetic field strength is varied ($B_0=[0.4, 0.8, 1.2, 1.6]\,\text{G}$). As can be seen, the stronger the magnetic field, the closer the position of the minimum magnetic energy to the FR. This produces a strong difference in the early stages of the FR evolution and in their velocities. For the later evolution all cases seem to be channeled in a same path. It must be noted that the CHs of these cases have field lines of the same shape, as discussed below. 
    \item   Cases 2, 5, 6 and 7 are shown in the middle panels of Fig.~\ref{F-Energy}. These CH configurations have different distances from the FR ($D=[150, 180, 250, 350]\,\text{Mm}$). For these cases we can see that the closer the CH to the FR the closer the position of the MME to the FR too. In addition, apparently the lower (in $y$-direction) the position of the minimum magnetic energy region, the slower the rising of the flux rope.   
    \item  The bottom panels of Fig.~\ref{F-Energy} show the paths of cases 8, 2, 9 and 10, where the simulated coronal holes have different widths ($W=[300, 400, 500, 600]\,\text{Mm}$). The wider the CH, the closer the position of the MME to the FR. For these cases, as in the previous item, the lower (in $y-$direction) the position of the minimum magnetic energy region, the slower rising of the flux rope.
\end{itemize} 
 For this qualitative analysis we note that the distance from the point of minimum magnetic energy to the flux rope (measured at $t=0\,$s) seems to play an important role mainly in the first stage of the evolution.

\begin{figure}    
   \centerline{\includegraphics[width=0.45\textwidth]{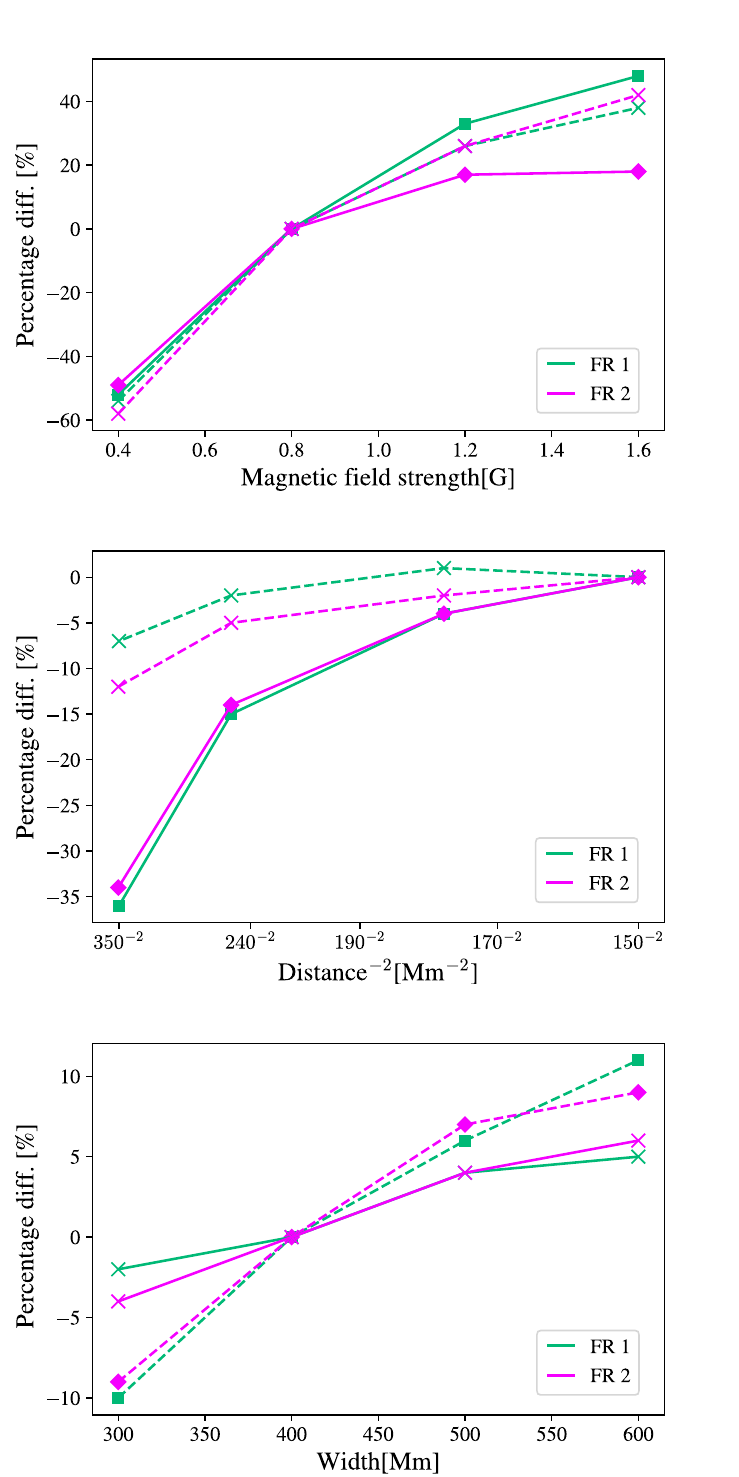}
              }
              \caption{ Percentage of $(d_0-d)/d_0 \,[\%]$ (dashed lines) and percentage of $(v-v_0)/v_0\,[\%]$ (solid lines) in function of the CH parameters. }
   \label{F-relation}
\end{figure}

In order to evaluate the relation between the position of the minimum magnetic energy region and the CME deflection, we compare the relative change in the initial distance between the MME point and the FR with respect to the reference case ($(d_0-d)/d_0 \,[\%]$) and the corresponding relative change in the deflection velocities $(v-v_0)/v_0\,[\%]$.
In other words, we want to know how much closer is the MME point to the FR for a given change in the deflection velocity.
In the top panel of Fig.~\ref{F-relation} we plot the relative changes of the initial MME distances and the deflection velocities for different magnetic field strengths.
It can be seen a trend where the stronger the magnetic field the closer the MME point to the FR and the larger the deflection, but the change in the deflection velocity can be greater or lesser than the change in the MME distance depending on the FR configuration.
In addition, it is observed that, for weak magnetic fields, there is a direct relationship between the relative change in the MME distance and the relative variation of the deflection velocity.
To consider now the effect of the distance from the flux rope to the coronal hole, we plot the middle panel of Fig.~\ref{F-relation}, where a similar trend to the previous case is observed: the closer the distance to the CH the closer the MME region to the FR and the larger the deflection.
However, in these results we can see that the relative changes in the deflection velocities is quite greater than the relative variation of the MME distance for both FR configurations.
The influence of the width of the coronal hole is presented in the bottom panel of Fig.~\ref{F-relation}. Here we see again a similar trend that, for this case, implies that the wider the CH the closer the MME point to the FR and the larger the deflection.
Contrarily to the results of the effect of distance to the CH analysis, in this case the relative change in the MME distance is larger than the relative variation of the deflection velocity.

\begin{figure}    
   \centerline{\includegraphics[width=0.45\textwidth]{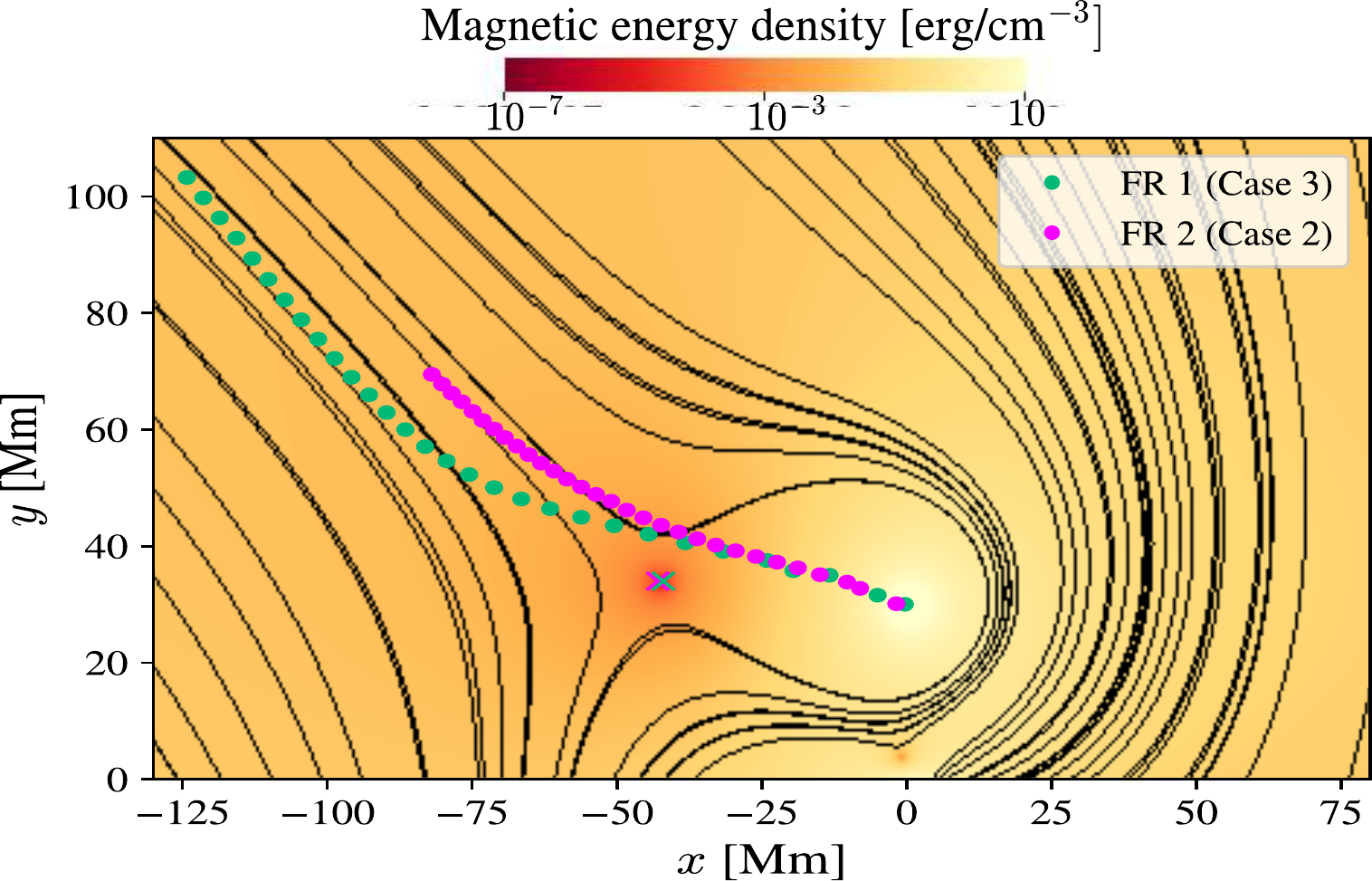}
              }
              \caption{ Path of the FR center (dots), and the position of the MME at $t=0\,$s (x), for different FRs and different magnetic field strengths of CH. The color map represents the magnetic energy density and the black lines represent the magnetic field lines. }
   \label{F-difdef}
\end{figure}

Considering the previous analysis, we can conclude that the closer de position of the MME point to the FR the larger the deflection of the CME, but this relation seems to be quite complex and depends on the other parameters of the problem.
This means that different configurations with equal initial distance between the MME point and the FR center will not necessarily exhibit similar deflections because other parameters strongly affect the CME trajectory, as shown in Fig.~{\ref{F-difdef}}.

In reference to the physical mechanisms involved in the CME deflections by the presence of coronal holes we identify the following aspects.
Firstly, we note that the MME point produced by the overlapping of the different magnetic structures is initially an attracting point to the flux rope, whose effect is more important in the early evolution of the ejection.
The reason by which the FR does not reach the MME point is because this minimum of magnetic energy is destroyed when the bow shock generated by the FR rising impacts it.
Therefore, although the FR is initially attracted by the MME region, it is deflected by another factor when the magnetic energy is homogenized. On the contrary, the FR should follow an inertial path. Based on the results showed in the first panel of Fig.~\ref{F-Energy}, we think that the second mechanism that triggers the deflection is the channeling of the FR, which is evident in Fig.~\ref{F-difdef}. The channel where the FR continues its trajectory is a product of the magnetic field lines of the CH in combination with the background magnetic field.
From our simulations, we understand that the coronal hole acts as a magnetic wall, which can have a remote repulsive action, through the formation of a MME and a channel that triggers the deflection of the flux rope. 

\section{Conclusions and discussion}
The study of CME deflections is of great interest to the space weather forecast. Many efforts have been carried out to understand how the intrinsic properties of the CME or the magnetic structures of the environment can influence the deflections. One of the more studied coronal magnetic structures are the coronal holes. To quantify the influence of CHs on the CME deflections, several authors use an ``influence parameter'' which depends on the intrinsic features of the coronal holes \citep{2006AdSpR..38..461C,2009JGRA..114.0A22G,2012JGRA..117.1103M}.
In our analysis of the early stages of a CME deflection in presence of a CH, we obtained that the deflection increases for wider CHs, with stronger magnetic fields, and decreases when the CH moves away from the flux rope.  Also, we found similar behaviors for two different FRs which suggest that the influence of the parameters of the CH are mostly independent of the FRs scenarios.

From the complete set of simulations a tendency is revealed: the deviation angle grows almost linearly with a  low deceleration. This deceleration could be attributed to the increasing distance between the CH and the CME as it rises which consequently reduces the CH influence in the CME trajectory. Moreover, we found a dimensionless parameter related with the velocity of the deflection that behaves similarly to the ``influence parameter'' proposed by \citet{2009JGRA..114.0A22G}. We obtained that the relative deflection velocity increases linearly with the magnetic field strength and decreases with the inverse square of the distance. However, the relative deflection velocities obtained from the width variations are lower than the values obtained from the proposed linear behavior. It is worth mentioning that such overestimation could be explained either by the inaccuracy of the proposed model as well as by an inaccurate definition of the CH width. The mathematical width used could be overestimating the effective size of the coronal hole.

 On the physical insight of the influence of coronal holes on CME deviations, we found some interesting results. We showed that a minimum magnetic energy (MME) point is produced by the overlapping of different magnetic structures, which changes the trajectory of the flux rope, mainly in its early evolution. There is a relationship between the speed of deviation of the CME and the initial distance between the point of MME and the FR; this relationship shows that both quantities depend on the parameters of the CH.  For the later evolution of the ejection, we notice a channeling of the flux rope. We proposed that the formation of the MME point and the channel that lead to the deflection of the FR, are a consequence of the presence of the CH, attributing to it the remote action of a magnetic wall.

To conclude, we reinforce the fact that the presence of coronal holes in the area of the CME formation is of crucial importance in the later evolution of the event. This systematic study of the influence of the coronal hole properties in the CME early evolution can help to characterize the effect of these magnetic structures in the full evolution of coronal mass ejections.

\acknowledgments
We would like to thank A. Costa for many stimulating discussions throughout this work. AS is doctoral fellow of CONICET. MC y GK are members of the Carrera del Investigador Cient\'ifico (CONICET). AS, MC y GK acknowledge support from ANPCyT under grant number PICT No. 2016-2480. AS y MC also acknowledge support by SECYT-UNC grant number PC No. 33620180101147CB. Also, we thank the Centro de C\'omputo de Alto Desempe\~no (UNC), where the simulations were carried out.

\bibliography{biblio}{}

\begin{thebibliography}{}
\expandafter\ifx\csname natexlab\endcsname\relax\def\natexlab#1{#1}\fi
\providecommand{\url}[1]{\href{#1}{#1}}
\providecommand{\dodoi}[1]{doi:~\href{http://doi.org/#1}{\nolinkurl{#1}}}
\providecommand{\doeprint}[1]{\href{http://ascl.net/#1}{\nolinkurl{http://ascl.net/#1}}}
\providecommand{\doarXiv}[1]{\href{https://arxiv.org/abs/#1}{\nolinkurl{https://arxiv.org/abs/#1}}}

\bibitem[{{Berger}(2014)}]{2014IAUS..300...15B}
{Berger}, T. 2014, in IAU Symposium, Vol. 300, Nature of Prominences and their
  Role in Space Weather, ed. B.~{Schmieder}, J.-M. {Malherbe}, \& S.~T. {Wu},
  15--29, \dodoi{10.1017/S1743921313010697}

\bibitem[{{Capannolo} {et~al.}(2017){Capannolo}, {Opher}, {Kay}, \& {Land
  i}}]{2017ApJ...839...37C}
{Capannolo}, L., {Opher}, M., {Kay}, C., \& {Land i}, E. 2017, \apj, 839, 37,
  \dodoi{10.3847/1538-4357/aa6a16}

\bibitem[{{C{\'e}cere} {et~al.}(2019){C{\'e}cere}, {Sieyra}, {Cremades},
  {Mierla}, {Sahade}, {Stenborg}, {Costa}, {West}, \&
  {D'Huys}}]{2019AdSpR..XX.XXXXC}
{C{\'e}cere}, M., {Sieyra}, M., {Cremades}, H., {et~al.} 2019, Advances in
  Space Research, 1, \dodoi{10.1016/j.asr.2019.08.043}

\bibitem[{{Cheng} {et~al.}(2012){Cheng}, {Zhang}, {Saar}, \&
  {Ding}}]{2012ApJ...761...62C}
{Cheng}, X., {Zhang}, J., {Saar}, S.~H., \& {Ding}, M.~D. 2012, \apj, 761, 62,
  \dodoi{10.1088/0004-637X/761/1/62}

\bibitem[{{Cremades} {et~al.}(2006){Cremades}, {Bothmer}, \&
  {Tripathi}}]{2006AdSpR..38..461C}
{Cremades}, H., {Bothmer}, V., \& {Tripathi}, D. 2006, Advances in Space
  Research, 38, 461, \dodoi{10.1016/j.asr.2005.01.095}

\bibitem[{{Forbes}(1990)}]{1990JGR....9511919F}
{Forbes}, T.~G. 1990, Journal of Geophysical Research, 95, 11919,
  \dodoi{10.1029/JA095iA08p11919}

\bibitem[{{Fryxell} {et~al.}(2000){Fryxell}, {Olson}, {Ricker}, {Timmes},
  {Zingale}, {Lamb}, {MacNeice}, {Rosner}, {Truran}, \&
  {Tufo}}]{2000ApJS..131..273F}
{Fryxell}, B., {Olson}, K., {Ricker}, P., {et~al.} 2000, The Astrophysical
  Journal Supplement Series, 131, 273, \dodoi{10.1086/317361}

\bibitem[{{Gopalswamy} {et~al.}(2009){Gopalswamy}, {M{\"a}kel{\"a}}, {Xie},
  {Akiyama}, \& {Yashiro}}]{2009JGRA..114.0A22G}
{Gopalswamy}, N., {M{\"a}kel{\"a}}, P., {Xie}, H., {Akiyama}, S., \& {Yashiro},
  S. 2009, Journal of Geophysical Research (Space Physics), 114, A00A22,
  \dodoi{10.1029/2008JA013686}

\bibitem[{{Heinemann} {et~al.}(2019){Heinemann}, {Temmer}, {Heinemann},
  {Dissauer}, {Samara}, {Jer{\v{c}}i{\'c}}, {Hofmeister}, \&
  {Veronig}}]{2019SoPh..294..144H}
{Heinemann}, S.~G., {Temmer}, M., {Heinemann}, N., {et~al.} 2019, \solphys,
  294, 144, \dodoi{10.1007/s11207-019-1539-y}

\bibitem[{{Hofmeister} {et~al.}(2017){Hofmeister}, {Veronig}, {Reiss},
  {Temmer}, {Vennerstrom}, {Vr{\v{s}}nak}, \& {Heber}}]{2017ApJ...835..268H}
{Hofmeister}, S.~J., {Veronig}, A., {Reiss}, M.~A., {et~al.} 2017, \apj, 835,
  268, \dodoi{10.3847/1538-4357/835/2/268}

\bibitem[{{Jin} {et~al.}(2017){Jin}, {Manchester}, {van der Holst}, {Sokolov},
  {T{\'o}th}, {Vourlidas}, {de Koning}, \& {Gombosi}}]{2017ApJ...834..172J}
{Jin}, M., {Manchester}, W.~B., {van der Holst}, B., {et~al.} 2017, \apj, 834,
  172, \dodoi{10.3847/1538-4357/834/2/172}

\bibitem[{{Kay} {et~al.}(2013){Kay}, {Opher}, \& {Evans}}]{2013ApJ...775....5K}
{Kay}, C., {Opher}, M., \& {Evans}, R.~M. 2013, \apj, 775, 5,
  \dodoi{10.1088/0004-637X/775/1/5}

\bibitem[{{Kay} {et~al.}(2015){Kay}, {Opher}, \& {Evans}}]{2015ApJ...805..168K}
---. 2015, \apj, 805, 168, \dodoi{10.1088/0004-637X/805/2/168}

\bibitem[{{Kilpua} {et~al.}(2009){Kilpua}, {Pomoell}, {Vourlidas}, {Vainio},
  {Luhmann}, {Li}, {Schroeder}, {Galvin}, \& {Simunac}}]{2009AnGeo..27.4491K}
{Kilpua}, E.~K.~J., {Pomoell}, J., {Vourlidas}, A., {et~al.} 2009, Annales
  Geophysicae, 27, 4491, \dodoi{10.5194/angeo-27-4491-2009}

\bibitem[{{Krause}(2019)}]{2019A&A...631A..68K}
{Krause}, G. 2019, \aap, 631, A68, \dodoi{10.1051/0004-6361/201936387}

\bibitem[{{Krause} {et~al.}(2018){Krause}, {C{\'e}cere}, {Zurbriggen}, {Costa},
  {Francile}, \& {Elaskar}}]{2018MNRAS.474..770K}
{Krause}, G., {C{\'e}cere}, M., {Zurbriggen}, E., {et~al.} 2018, \mnras, 474,
  770, \dodoi{10.1093/mnras/stx2817}

\bibitem[{{Landi} {et~al.}(2010){Landi}, {Raymond}, {Miralles}, \&
  {Hara}}]{2010ApJ...711...75L}
{Landi}, E., {Raymond}, J.~C., {Miralles}, M.~P., \& {Hara}, H. 2010, \apj,
  711, 75, \dodoi{10.1088/0004-637X/711/1/75}

\bibitem[{{Lee} \& {Deane}(2009)}]{2009JCoPh.228..952L}
{Lee}, D., \& {Deane}, A.~E. 2009, Journal of Computational Physics, 228, 952,
  \dodoi{10.1016/j.jcp.2008.08.026}

\bibitem[{{Liewer} {et~al.}(2015){Liewer}, {Panasenco}, {Vourlidas}, \&
  {Colaninno}}]{2015SoPh..290.3343L}
{Liewer}, P., {Panasenco}, O., {Vourlidas}, A., \& {Colaninno}, R. 2015,
  \solphys, 290, 3343, \dodoi{10.1007/s11207-015-0794-9}

\bibitem[{{Lugaz} {et~al.}(2011){Lugaz}, {Downs}, {Shibata}, {Roussev}, {Asai},
  \& {Gombosi}}]{2011ApJ...738..127L}
{Lugaz}, N., {Downs}, C., {Shibata}, K., {et~al.} 2011, \apj, 738, 127,
  \dodoi{10.1088/0004-637X/738/2/127}

\bibitem[{{Lynch} \& {Edmondson}(2013)}]{2013ApJ...764...87L}
{Lynch}, B.~J., \& {Edmondson}, J.~K. 2013, \apj, 764, 87,
  \dodoi{10.1088/0004-637X/764/1/87}

\bibitem[{{M{\"a}kel{\"a}} {et~al.}(2013){M{\"a}kel{\"a}}, {Gopalswamy}, {Xie},
  {Mohamed}, {Akiyama}, \& {Yashiro}}]{2013SoPh..284...59M}
{M{\"a}kel{\"a}}, P., {Gopalswamy}, N., {Xie}, H., {et~al.} 2013, \solphys,
  284, 59, \dodoi{10.1007/s11207-012-0211-6}

\bibitem[{{Mei} {et~al.}(2012){Mei}, {Shen}, {Wu}, {Lin}, {Murphy}, \&
  {Roussev}}]{2012MNRAS.425.2824M}
{Mei}, Z., {Shen}, C., {Wu}, N., {et~al.} 2012, \mnras, 425, 2824,
  \dodoi{10.1111/j.1365-2966.2012.21625.x}

\bibitem[{{Mohamed} {et~al.}(2012){Mohamed}, {Gopalswamy}, {Yashiro},
  {Akiyama}, {M{\"a}kel{\"a}}, {Xie}, \& {Jung}}]{2012JGRA..117.1103M}
{Mohamed}, A.~A., {Gopalswamy}, N., {Yashiro}, S., {et~al.} 2012, Journal of
  Geophysical Research (Space Physics), 117, A01103,
  \dodoi{10.1029/2011JA016589}

\bibitem[{{M{\"o}stl} {et~al.}(2015){M{\"o}stl}, {Rollett}, {Frahm}, {Liu},
  {Long}, {Colaninno}, {Reiss}, {Temmer}, {Farrugia}, {Posner}, {Dumbovi{\'c}},
  {Janvier}, {D{\'e}moulin}, {Boakes}, {Devos}, {Kraaikamp}, {Mays}, \& {Vr{\v
  s}nak}}]{2015NatCo...6E7135M}
{M{\"o}stl}, C., {Rollett}, T., {Frahm}, R.~A., {et~al.} 2015, Nature
  Communications, 6, 7135, \dodoi{10.1038/ncomms8135}

\bibitem[{{Panasenco} {et~al.}(2011){Panasenco}, {Martin}, {Joshi}, \&
  {Srivastava}}]{2011JASTP..73.1129P}
{Panasenco}, O., {Martin}, S., {Joshi}, A.~D., \& {Srivastava}, N. 2011,
  Journal of Atmospheric and Solar-Terrestrial Physics, 73, 1129,
  \dodoi{10.1016/j.jastp.2010.09.010}

\bibitem[{{Panasenco} {et~al.}(2013){Panasenco}, {Martin}, {Velli}, \&
  {Vourlidas}}]{2013SoPh..287..391P}
{Panasenco}, O., {Martin}, S.~F., {Velli}, M., \& {Vourlidas}, A. 2013,
  \solphys, 287, 391, \dodoi{10.1007/s11207-012-0194-3}

\bibitem[{{Pascoe} {et~al.}(2014){Pascoe}, {Nakariakov}, \&
  {Kupriyanova}}]{2014A&A...568A..20P}
{Pascoe}, D.~J., {Nakariakov}, V.~M., \& {Kupriyanova}, E.~G. 2014, \aap, 568,
  A20, \dodoi{10.1051/0004-6361/201423931}

\bibitem[{{Robertson} \& {Priest}(1987)}]{1987SoPh..114..311R}
{Robertson}, J.~A., \& {Priest}, E.~R. 1987, \solphys, 114, 311,
  \dodoi{10.1007/BF00167348}

\bibitem[{{Syntelis} {et~al.}(2016){Syntelis}, {Gontikakis}, {Patsourakos}, \&
  {Tsinganos}}]{2016A&A...588A..16S}
{Syntelis}, P., {Gontikakis}, C., {Patsourakos}, S., \& {Tsinganos}, K. 2016,
  \aap, 588, A16, \dodoi{10.1051/0004-6361/201526829}

\bibitem[{{van der Holst} {et~al.}(2010){van der Holst}, {Manchester},
  {Frazin}, {V{\'a}squez}, {T{\'o}th}, \& {Gombosi}}]{2010ApJ...725.1373V}
{van der Holst}, B., {Manchester}, W.~B., I., {Frazin}, R.~A., {et~al.} 2010,
  \apj, 725, 1373, \dodoi{10.1088/0004-637X/725/1/1373}

\bibitem[{{van Tend} \& {Kuperus}(1978)}]{1978SoPh...59..115V}
{van Tend}, W., \& {Kuperus}, M. 1978, \solphys, 59, 115,
  \dodoi{10.1007/BF00154935}

\bibitem[{{V{\'a}squez}(2016)}]{2016AdSpR..57.1286V}
{V{\'a}squez}, A.~M. 2016, Advances in Space Research, 57, 1286,
  \dodoi{10.1016/j.asr.2015.05.047}

\bibitem[{{Wang} {et~al.}(2019){Wang}, {Hoeksema}, \&
  {Liu}}]{2019arXiv190906410W}
{Wang}, J., {Hoeksema}, J.~T., \& {Liu}, S. 2019, arXiv e-prints,
  arXiv:1909.06410.
\newblock \doarXiv{1909.06410}

\bibitem[{{Xie} {et~al.}(2009){Xie}, {St.~Cyr}, {Gopalswamy}, {Yashiro},
  {Krall}, {Kramar}, \& {Davila}}]{2009SoPh..259..143X}
{Xie}, H., {St.~Cyr}, O.~C., {Gopalswamy}, N., {et~al.} 2009, \solphys, 259,
  143, \dodoi{10.1007/s11207-009-9422-x}

\bibitem[{{Yang} {et~al.}(2018){Yang}, {Dai}, {Chen}, {Li}, \&
  {Jiang}}]{2018ApJ...862...86Y}
{Yang}, J., {Dai}, J., {Chen}, H., {Li}, H., \& {Jiang}, Y. 2018, \apj, 862,
  86, \dodoi{10.3847/1538-4357/aaccfd}

\bibitem[{{Zhou} {et~al.}(2014){Zhou}, {Feng}, \& {Zhao}}]{2014JGRA..119.9321Z}
{Zhou}, Y., {Feng}, X., \& {Zhao}, X. 2014, Journal of Geophysical Research
  (Space Physics), 119, 9321, \dodoi{10.1002/2014JA020347}

\bibitem[{{Zhou} \& {Feng}(2013)}]{2013JGRA..118.6007Z}
{Zhou}, Y.~F., \& {Feng}, X.~S. 2013, Journal of Geophysical Research (Space
  Physics), 118, 6007, \dodoi{10.1002/2013JA018976}

\bibitem[{{Zhuang} {et~al.}(2017){Zhuang}, {Wang}, {Shen}, {Liu}, {Wang},
  {Pan}, {Li}, \& {Liu}}]{2017ApJ...845..117Z}
{Zhuang}, B., {Wang}, Y., {Shen}, C., {et~al.} 2017, \apj, 845, 117,
  \dodoi{10.3847/1538-4357/aa7fc0}

\bibitem[{{Zuccarello} {et~al.}(2012){Zuccarello}, {Bemporad}, {Jacobs},
  {Mierla}, {Poedts}, \& {Zuccarello}}]{2012ApJ...744...66Z}
{Zuccarello}, F.~P., {Bemporad}, A., {Jacobs}, C., {et~al.} 2012, \apj, 744,
  66, \dodoi{10.1088/0004-637X/744/1/66}

\end{thebibliography}
\bibliographystyle{aasjournal}



\end{document}